# Self-Assembly on a Cylinder: A Model System for Understanding the Constraint of Commensurability

D. A. Wood, C. D. Santangelo, A. D. Dinsmore*
*June, 2013.*

A crystal lattice, when confined to the surface of a cylinder, must have a periodic structure that is commensurate with the cylinder circumference. This constraint can frustrate the system, leading to oblique crystal lattices or to structures with a chiral seam known as a 'line slip' phase, neither of which are stable for isotropic particles in equilibrium on flat surfaces. In this study, we use molecular dynamics simulations to find the steady-state structure of spherical particles with short-range repulsion and long-range attraction far below the melting temperature. We vary the range of attraction using the Lennard-Jones and Morse potentials and find that a shorter-range attraction favors the line-slip. We develop a simple model based only on geometry and bond energy to predict when the crystal or line-slip phases should appear, and find reasonable agreement with the simulations. The simplicity of this model allows us to understand the influence of the commensurability constraint, an understanding that might be extended into the more general problem of self-assembling particles in strongly confined spaces.

## INTRODUCTION

A two dimensional crystal confined to a curved surface exhibits a variety of interesting behaviors that are quite different from the behaviors that arise when it is confined to a planar surface. These differences are a product of the curvature and topology of the surface itself and appear even at zero temperature [1]. A useful measure for describing such phenomena is Gaussian curvature–defined as the inverse product of the two principal radii of curvature. When constrained to a surface with non-zero Gaussian curvature, such as a sphere or a torus, hexagonal crystals form isolated 5-fold disclinations or dislocation-disclination combinations known as scars [2, 3]. These defects play an important role in of self-assembly on surfaces with non-zero Gaussian curvature [4], such as in the assembly of viral capsids and other biological structures [5, 6].

On surfaces of zero Gaussian curvature, no particular requirement to form such defects exists because a two dimensional crystal can lie on its surface without in-plane stretching. If, however, we were to confine the crystal to a closed surface, the finite size of the system could induce defects in the crystal [7], the properties of which are distinct from those induced by Gaussian curvature [1]. A cylinder provides a simple example of exactly such a surface: it has vanishing Gaussian curvature and a finite-size constraint arising from periodicity. Indeed, the requirement that a crystal lattice be invariant under a full axial rotation (i.e., that it be commensurate with the substrate circumference) leads to a variety of crystalline and non-crystalline structures that are distinct from those found on planar surfaces. The potential for surface shape and size to tune in-plane packing may be relevant in biological or technological examples of self-assembly such as coating the surface of fibers, the arrangement of kernels on a corncob [8], assembly of proteins on membrane tubules [9], growth of bacterial cell walls [10], proteins or DNA on microtubules [11], surfactants on a nanometer-scale cylinder [12], or proteins around RNA as in helical viral capsids [13].

Previous works have used computer simulations to find the densest packing of hard spheres *inside* a cylinder; these studies found that uniform hexagonal crystals are found only for a discrete set of cylinder radii, between which the spheres form structures with a chiral seam (referred to as a 'line-slip' by Mughal et al.) [14-17]. Experimentally, interior packing has been investigated through soft colloidal spheres confined to cylindrical channels [18] or rigid colloid spheres in rectilinear channels [19], and fullerene nanospheres confined within carbon nanotubes [20].

Confinement to the *surface* of a cylinder, on the other hand, has received comparatively little attention. Mughal et. al make a link between surface packing and interior packing, and predict the maximal packing structures formed by hard spheres on a cylinder [15, 16]. However, a maximally packed structure is akin to a system of hard-sphere particles at infinite pressure, which is of limited use in predicting self-assembly. Other authors report on assembly of purely repulsive particles [21-23] or more complex systems with competing species of oppositely charged particles [11, 24]. While these studies demonstrate a wide array of possible phases, they leave open the question of assembly of particles with a common class of interaction potential.

In this paper, we examine spherical particles constrained to the surface of a cylinder as a model system for understanding how frustration created by the incommensurability of the preferred packing with the available area can influence the stable solid structures found there. In particular, we investigate the role of an isotropic interaction that has a steep, short-ranged repulsion and a longer-ranged attraction. We vary the range of the attraction and apply simulated annealing to probe the range of stable structures that arise at temperatures well below the melting point. We find that a finite interaction (Lennard-Jones) substantially broadens the range of cylinder radii over which stable and uniform crystals are found compared to hard spheres. In such cases, we find that that the cylinder stabilizes an oblique lattice structure, which is not found in equilibrium on a planar surface. We also find finite ranges of cylinder size that induce "line-slip phases," characterized by a helical defect that separates two regions with the same crystal lattice and orientation. The line-slip phase resembles structures previously found with hard spheres [15, 16]. When the range of attraction is decreased relative to the sphere size, the area in parameter space over which stable and uniform crystals are found decreases and the line-slip structures become more prevalent. We find that these behaviors can be predicted with surprising accuracy using a simple one-dimensional model. The simplicity of this model allows us to understand the basic mechanism at work in this geometry — an understanding which might be extended into the more general problem of self-assembling particles in strongly confined spaces.

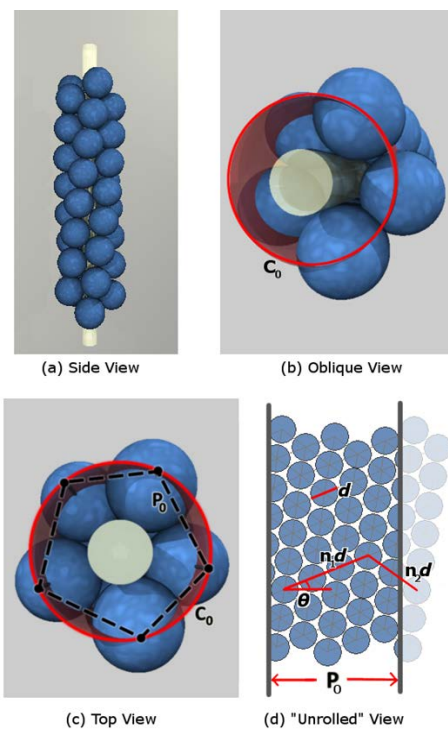

**Fig. 1.** Illustrations of the particle-and-cylinder system. (a) Spheres are attached to the cylinder surface in a perfect hexagonal lattice. (b) The same lattice shown at an oblique angle to illustrate the difference between the inner cylinder circumference (which belongs to the physical rod) and the outer cylinder circumference $C_0$ (which intersects the centers of mass of the particles). (c) A top view of this lattice, projected onto a plane perpendicular to the cylinder axis. Here, $C_0$ is the circumference of the circle that intersects the centers of the particles, and $P_0$ is the perimeter of the corresponding inscribed polyhedron. (d) The same lattice, projected into a plane created by slicing the crystal vertically and laying it on a flat surface. Note that this is not the plane defined by the surface of $C_0$ (which would distort the metric). The angle θ and the lattice constants $n_1$ and $n_2$ have been labeled.

**Defining the Geometry and Commensurability Constraint.**

We begin by considering the example of a *perfect hexagonal* lattice that is wrapped onto the surface of a cylinder (Fig. 1). This example provides a useful reference for our later discussion of the structures found in simulations. In this problem, the fact that the Gaussian curvature is zero everywhere avoids curvature-induced defects. Instead, the structure is subject to the constraint that it be invariant under a $2\pi$ rotation about the cylinder axis. This discrete rotational symmetry leads directly to the conclusion that only a discrete set of cylinder circumferences can accommodate a perfect hexagonal lattice. Below, we refer to this discrete set of configurations with the subscript 0.

To define the geometry, we consider the cylinder of circumference $C$, whose surface contains the center of mass of each particle (rather than the cylinder whose surface lies tangent to each sphere). The unit vector $\hat{z}$ lies along the cylinder axis and $\hat{\varphi}$ lies in the circumferential direction. To express the $2\pi$ rotation symmetry, we first define two lattice translation vectors $a_1$ and $a_2$. For convenience, we further define $a_1$ as the lattice vector that lies closest to the $\hat{\varphi}$ direction. Because we are interested in the packing of three-dimensional spheres, we define $a_{1,2}$ as the nearest-neighbor spacing in three-dimensional Euclidean space. Perhaps the most straightforward way to write the rotational symmetry condition is to require that the pathway defined by $n_1$ steps along $a_1$ and $n_2$ steps along $a_2$ form a simple polygon enclosing the cylinder. The projection of the perimeter of this polygon along $\hat{\varphi}$ (Fig. 1c) is a useful parameter for defining the commensurability constraint. We label this projected polygon $P$.

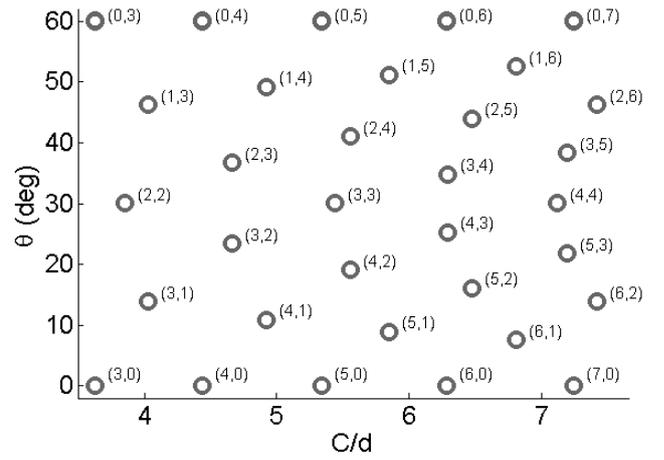

**Fig. 2.** The values of $C_0(n_1,n_2)$ are plotted alongside their corresponding set of integers $(n_1,n_2)$. These points mark locations where a defect-free hexagonal crystal fits on the cylinder surface. Exchanging $n_1$ and $n_2$ results in a reflection across θ=30°. The two values correspond to equivalent configurations (a chiral pair)

For the case of a perfect hexagonal crystal, $|a_{1,2}| = d$ (the particle diameter) and $a_{1,2}$ are separated by an angle of $\pi/6$. The $2\pi$-rotational symmetry then reduces to the requirement that the perimeter of the polygon be given by $P_0$, which is defined by [8, 21, 25, 26]

$$P_0(n_1,n_2) = d\,(n_1^2 + n_2^2 + n_1 n_2)^{1/2}. \qquad \text{(Eq. 1)}$$

Likewise, maintaining perfect hexagonal symmetry also requires the circumference $C$ (Fig. 1c) to be expressible in terms of $n_1$ and $n_2$. This relationship may be found from basic geometry, and the result is a transcendental equation that defines the allowed circumferences $C_0(n_1,n_2)$ for a hexagonal lattice:

$$n_1 \cdot \sin^{-1}\left[\left(\frac{\pi d}{2C_0}\right) \cdot \frac{2n_1+n_2}{\sqrt{n_1^2+n_1 n_2+n_2^2}}\right] + n_2 \cdot \sin^{-1}\left[\left(\frac{\pi d}{2C_0}\right) \cdot \frac{2n_2+n_1}{\sqrt{n_1^2+n_1 n_2+n_2^2}}\right] = \pi. \qquad \text{(Eq. 2)}$$

For each $(n_1,n_2)$, there is also a fixed orientation of the lattice, $\theta_0$, which we will define as the angle between $a_1$ and $\hat{\varphi}$ (Fig. 1d):

$$\theta_0(n_1,n_2) = \tan^{-1}\left(\frac{\sqrt{3}\,n_2}{2n_1+n_2}\right). \qquad \text{(Eq. 3)}$$

For each value of $C_0$, there are in general two distinct values of $\theta_0$, which correspond to permuted values of $n_1$ and $n_2$. These two structures have opposite chirality and are physically indistinguishable.

In Figure 2, the open circles show $(C_0(n_1,n_2), \theta_0(n_1,n_2))$: these are the configurations that are allowed for perfect hexagonal lattices as defined by equations (2,3). Each point is labeled by the corresponding set of integers $(n_1,n_2)$. Note that configurations with $\theta_0 \geq 60°$ are identical to configurations with $0 \leq \theta_0 < 60°$ because of the 6-fold rotational symmetry of the lattice.

The question that we now address with simulations is how the set of configurations is altered when there is a finite interaction potential and the spheres can move freely along the cylinder's surface. Or, more to the point, what structures appear when $C \neq C_0$?

# SIMULATIONS

We employed computer simulations to find the steady state structures that appear under conditions of different interaction potentials and across a range of cylinder circumferences. In particular, two different potentials were used in our simulations: the Lennard-Jones potential, which we define as

$$V_{LJ}(r) = \epsilon\left[\left(\frac{d}{r}\right)^{12} - 2\left(\frac{d}{r}\right)^{6}\right],$$  (Eq. 4)

and the Morse potential, which is defined as

$$V_{Morse}(r) = \epsilon\left[e^{-2\gamma d\left(\frac{r}{d}-1\right)} - 2e^{-\gamma d\left(\frac{r}{d}-1\right)}\right],$$  (Eq. 5)

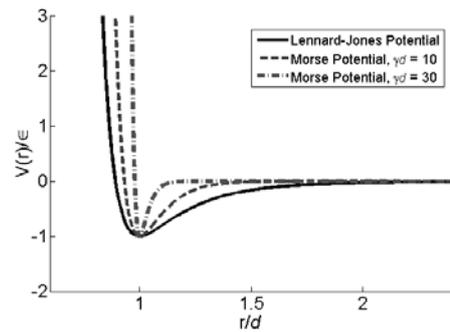

**Fig. 3.** The pairwise potential as a function of particle separation for both Lennard-Jones and Morse potentials. The γd=10 and γd=30 Morse potentials have a shorter ranged attraction and a steeper core repulsion than the Lennard-Jones potential.

where ϵ controls the interaction strength and the dimensionless parameter γd controls the range of attraction (a large γd corresponds to a short-range potential). In both cases, d represents the location of the potential minima and is distinct from the traditional hard-core separation distance. For reference, the Lennard-Jones potential has been plotted in Fig. 3 alongside the Morse potential for γd = 10 and γd = 30, the two values used in our simulations.

Although this problem is three-dimensional, the requirement that the particles lie on the surface of a cylinder allows us to translate it into a two-dimensional equivalent. More precisely, we represent the cylinder's surface in planar Cartesian space by letting $[C\varphi/2\pi, z] \rightarrow [x, y]$. In this representation, the x-axis necessarily has periodic boundary conditions and the y-axis does not.

The chosen interaction potential must also be adjusted as part of this two-dimensional mapping; interactions between particles must depend on their separation in Euclidean space, whereas the simulation coordinates are measured along the surface of the cylinder (Fig. 1c). This difference leads to a distortion along the $\hat{x}$ (or $\hat{\varphi}$) direction, which we account for in the simulations. For any two particles separated by $\Delta x$ and $\Delta y$ on the planar surface, the true Euclidean separation, $r$, is computed via

$$r = \sqrt{(C/\pi)^2 \cdot \sin^2(\Delta x \cdot \pi/C) + \Delta y^2}.$$  (Eq. 6)

For these two particles, the magnitude of the interaction force, $|\vec{F}|$, is obtained from the force-distance curves appropriate for the Lennard-Jones or Morse cases. The component of the force along the circumferential axis ($\hat{x}$) is computed as

$$F_x = |\vec{F}| \cdot (C/2\pi r) \cdot \sin(\Delta x \cdot 2\pi/C),$$  (Eq. 7a)

and the force along the cylinder axis ($\hat{y}$) is computed as

$$F_y = |\vec{F}| \cdot (\Delta y/r).$$  (Eq. 7b)

Therefore, the 2D projection of the pairwise potentials onto the (x,y) plane is slightly anisotropic.

Our simulations were performed using the LAMMPS molecular dynamics software [http://lammps.sandia.gov] [27]. These simulations utilized Langevin dynamics to evolve the system, which allowed us to indirectly control the system's temperature over the course of each simulation (further details on this procedure can be found in [28]). In all cases, particles were cooled from $T_i=2\epsilon/k_B$ to $T_f=0.01\epsilon/k_B$ via simulated annealing over a period of $5\times10^5$ iterations. One distinct advantage of using simulated annealing as opposed to strictly minimizing energies or solving the equations analytically is that it allows us to capture intermediate metastable configurations; we return to this point near the end of the article.

Prior to the start of each simulation, the total number of particles was chosen such that the area fraction of the system was equal to 0.7; both the volume and number of particles in the system were then kept fixed

---

throughout the annealing process. Each iteration of the algorithm represented a time step of 0.02s, and computed only the interactions between pairs of particles within $2d$ of each other. The mean squared displacement of an individual particle during each time step was $(2^{-1/3} \times 10^{-2} k_B T/\epsilon)d^2$. At the beginning of this process, particles were randomized at high temperatures, then allowed to diffuse freely in a medium with an effective viscosity of $\eta=0.95$ mPa/s (similar to water at room temperature). Once the particles were cooled into a stable crystal and the annealing was complete, the coordinates of each sphere were recorded. This process was then repeated with a new value of $C$ and with a new arrangement of particles.

In order to check the validity of this method, a few simulations were instead cooled to $T_f=10^{-4}\epsilon/k_B$ over a period of $5 \times 10^7$ iterations, with no observed difference in results. We also performed several spatially large simulations, where the y-axis boundaries were separated by a distance of several hundred particle diameters. In these cases, the system was large enough to form crystalline domain boundaries, but we found the structure of each individual domain to be no different from those observed in smaller crystals at the same circumference. In other words, we found that the distance between the x-axis boundaries (the cylinder circumference) is the only important dimension of the system.

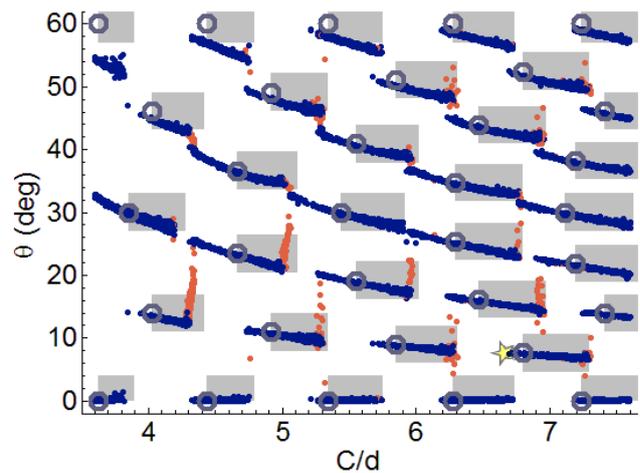

**Fig. 4.** Steady state configurations observed in computer simulations with Lennard-Jones interactions, plotted as $\theta$ vs $C/d$. Red/gray data points indicate the existence of a line-slip phase, and blue/black data points indicate a continuous uniform crystal. The large circles label the same analytically determined values for $C_0$ as in Fig. 2, and the gray blocks show defect-free regions as predicted by Eq. 9 and Eq. 11. The plot's axes span from $\theta=0$ to $\theta=60°$ because of the six-fold symmetry of the lattice. The star at $(C/d,\theta)=(6.76, 7.45°)$ indicates the location of the simulation examined in greater detail in Fig. 9.

The annealing process was completed using the interaction potentials of Eq. 4 and Eq. 5, and across several thousand cylinder circumferences between $C/d=3.6$ and $C/d=7.6$. For each of these simulations, we used the sphere positions to measure the average $\theta$ between the crystal lattice and the cylinder. This was accomplished through the following algorithm: first, every particle with six neighbors was indexed. Next, we recorded the angles defined by the $\hat{\varphi}$ axis and the bonds between each of these spheres and each of its neighbors. Using this data, we created a histogram of all angles, which necessarily has six distinct peaks. If two or more sets of six peaks were found, a polycrystalline structure was likely present (an inference confirmed by visual inspection of several such cases), and the data was rejected. The angles comprising the peak closest to 0° were then averaged. This value, which represents the angle of the lattice $\theta$, was then plotted as a function of $C/d$. Note, however, that this definition of $\theta$ measures the structure of the crystal in a way that is slightly different (though physically more meaningful) from the one used in Eq. 3, as described in the supplemental section [29]. As will be seen below, this leads to a very small discrepancy between our theory and our data that vanishes for large values of $C/d$.

**RESULTS**

We represent the steady-state configurations of the spheres by plots of $(C/d, \theta)$. This data, for systems of Lennard-Jones particles, is presented in Fig. 4. The pale grey bars in the background of this plot correspond to the theory derived in the next section. As expected, we find the uniform crystal phases (shown in blue/black) at the discrete values of $(C_0/d, \theta_0)$ associated with a perfect hexagonal lattice. We note, however, that for small $C/d$ the values of $\theta_0$ with $n_2 > n_1$ are slightly different from the values of $\theta$ extracted from our simulations. This is due to a subtle difference in the way the two angles are defined, which becomes small as $C/d$ increases [29].

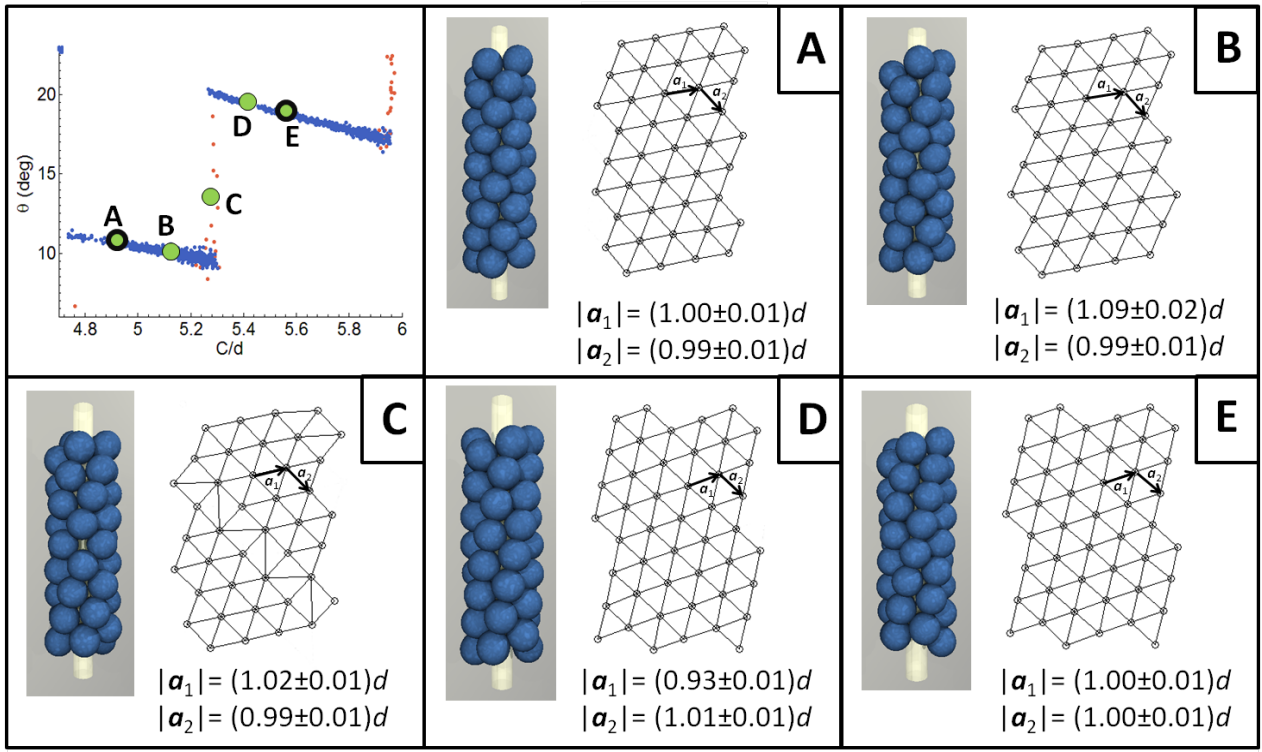

**Fig. 5.** Steady-state configurations obtained from the simulations are shown for a range of values of $C/d$ to better illustrate the structures found in Fig. 4. In each case, the average magnitude of the lattice vectors, $a_1$ and $a_2$, are provided alongside their standard deviation. The line segments show the Delaunay triangulation of the lattice, on top of which $a_1$ and $a_2$ have also been drawn. (A) A hexagonal lattice appears at the value of $C_0$ for $(n_1,n_2) = (4,1)$. (B) The lattice stretches along $a_1$ as $C/d$ increases. (C) A 'line slip' structure with a helical defect emerges in response to the strain placed on the crystal lattice. This is manifested in alternating pairs of 5-7 disclinations which span the length of the lattice. (D) The crystal, arranged at a new angle, is slightly compressed along $a_1$. (E) The lattice is again hexagonal at $C_0$ corresponding to $(n_1,n_2) = (4,2)$.

In contrast to the hard-sphere case, the uniform lattices persist across islands of parameter space surrounding each $(C_0/d, \theta_0)$. Between these islands of uniform lattices, we find narrow regions of parameter space that have structures with a chiral defect (orange/gray points); these correspond to the line-slip phases reported for hard spheres [16]. One (perhaps unexpected) feature of this graph is its asymmetry: an expansion of $C$ away from $C_0$ systematically leads to a reduction of $\theta$. This is due to the fact that, when a perfect hexagonal lattice cannot assemble on the cylinder at any angle, it becomes preferential to form an oblique lattice instead by stretching along the $\hat{\varphi}$ axis– a feature examined in greater detail in the following paragraphs. Chiral pairs in this state are no longer similar under rotation, but are instead similar only by reflection across $\theta = 0°$. The redundant data was omitted from the plot.

In Fig. 5, we show the real structure at selected points in the $(C,\theta)$ plane. Figure 5a shows the structure of a hexagonal lattice observed at the $(C_0, \theta_0)$ corresponding to $(n_1,n_2) = (4,1)$. As one increases the circumference of the cylinder away from some particular value of $C_0$, the preferred state becomes an oblique lattice, where the lattice unit vector nearest to the circumferential axis of the cylinder ($a_1$) is slightly increased and the other lattice vector ($a_2$) remains unchanged within uncertainty (Fig. 5b). Hence, we find that the cylinder stabilizes a new oblique crystal symmetry, which is not found on planar surfaces.

When the circumference is increased further, the strain proves to be too large and an oblique lattice at this orientation is no longer viable. The lattice then develops a chiral defect, which is the line-slip structure reported earlier [15]. This structure is shown in Fig. 5c. The line-slip defect itself consists of a line of 5-fold and 7-fold disclinations, as shown by the Delaunay triangulation of Fig. 5c. Although it seems to resemble a single dislocation line (of the sort that might be found in a polycrystalline structure on a planar surface), it actually consists of *two* distinct, adjacent dislocation lines. Here, the lines have zero net Burgers vector, separating two crystal phases that have the same symmetry and orientation.

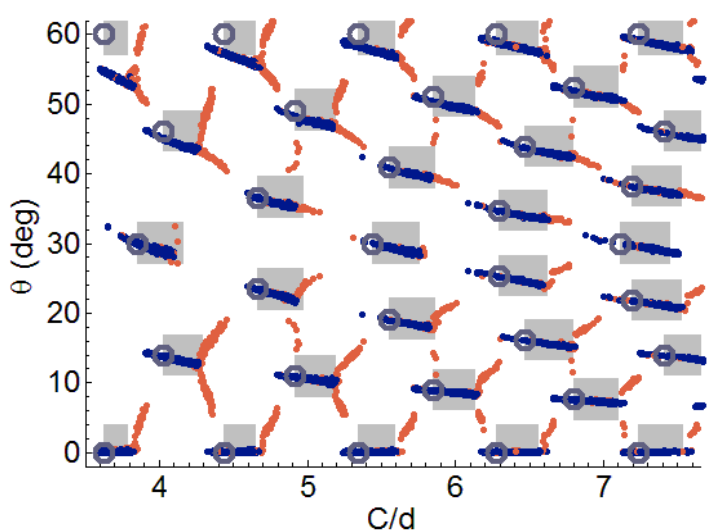

**Fig. 6.** Steady state configurations observed in computer simulations using the Morse potential with γd=10. Red/gray data points represent a line-slip phase; blue/black data points represent continuous uniform crystals. The large circles label the analytically determined vales for $C_0$ as in Fig. 2, and the gray blocks show defect-free regions as predicted by Eq. 9 and Eq. 11.

With still greater expansion of *C*, the observed structure shifts to a new crystal; this lattice is also slightly oblique, but is compressed along $a_1$ rather than stretched (Fig. 5d). This lattice corresponds to a distortion of the hexagonal lattice with $(n_1,n_2)$ = (4,2). As the circumference is increased, the lattice compression steadily decreases until the crystal is once again hexagonal (Fig 5e). The rest of the plot in Fig. 4 behaves in much the same manner as described above.

We emphasize that the transitions that occur between Fig. 5b and Fig. 5d are sharp. In fact, at no point do the lattices in Fig. 5b and Fig. 5d both appear simultaneously. Hence the line slip phase does not correspond to twinning, but instead appears to be a stable structure. Indeed, we found helical defects identical to the one in Fig. 5c in simulations with y-axis boundaries separated by several hundred particle diameters, as well as in simulations with annealing schedules several orders of magnitude longer than those shown here. This observation suggests that the line slip phase is a minimum-free-energy state that is stabilized by the frustration. We re-examine this idea in the next section, where we develop a model to compare the energies of the line-slip and crystal phases.

We now turn to the results obtained with interaction potentials with shorter-range attraction. These results allow us to separate the roles of the commensurability constraint from those of the interaction potential assigned to the particles. Figure 6 shows the structures observed with the Morse potential with γd = 10. The results are represented in the same (*C/d*, θ) plane as Fig. 4. As in the Lennard-Jones case, we find regions of uniform crystal that correspond, in general, to oblique distortions of the hexagonal lattice indexed by ($n_1$, $n_2$). Transitions between crystalline and line-slip structures are as sharp here as in the Lennard-Jones case. Unlike the Lennard-Jones case, however, we find that the line-slip structure supplants the crystal phase over a much broader region in parameter space. The reduced range of existence of a uniform crystal when *C* < $C_0$ might arise from the steeper repulsion of the Morse potential, as shown in Fig. 3. Similarly, the reduced range of uniform crystals when *C* > $C_0$ may be caused by the shorter range of attraction. We return to this point below in the context of our model. Also note that, for the case of *C* > $C_0$, the defect regions branch off of the islands of uniform crystals in opposite directions along the θ axis. This effect is caused by the additional chirality of the line-slip seam (see Fig. 5c), which is independent of the chirality of the lattice as a whole.

For the still shorter-range attraction, we find a continued trend toward increased prevalence of the line-slip structure. Figure 7 shows the observed structures for the Morse potential with γd = 30. Here, the line-slip states dominate the graph almost entirely and the uniform crystal phase is found only within narrow regions near ($C_0/d$, $θ_0$). In order to better explain this phenomenon, we now seek to develop a model that can predict where these transitions occur.

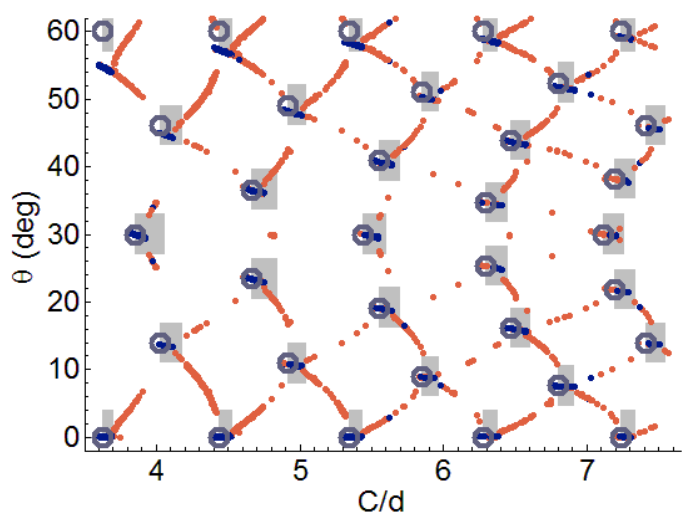

**Fig. 7.** Steady state configurations observed in computer simulations using the Morse potential with γd=30. Red/gray data points represent a line-slip phase, and blue/black data points represent continuous uniform crystals. Here, the line-slip phase predominates

## ONE-DIMENSIONAL MODEL

We propose a simplified model that succeeds in describing the major findings of the simulations. Let us first consider some $C_0/d$ for which the preferred lattice angle, $θ_0$, is 0° (i.e. $n_2 = 0$ and the particles are stacked in rings around the cylinder, as in Fig. 8c). To further simplify the model, we will consider only a single one-dimensional ring of particles along $\boldsymbol{a}_1$ and work in the low-temperature limit where entropy is not dominant. If the ring's circumference is increased beyond $C_0$ and the particles are then allowed to rearrange themselves, one can imagine two possible configurations emerging: (a) all the particles are spread uniformly along the ring and share the extra distance, or (b) the particles cluster and move all of the added strain to a single pair (Fig. 8). Ignoring all but nearest-neighbor interactions, we can write the total energy of these two configurations generically for some arbitrary short-range attractive potential V(r);

$$E_{uniform} = N \cdot V(P/N) \tag{Eq. 8a}$$

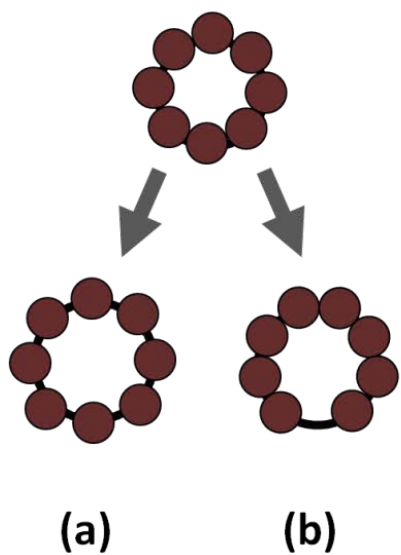 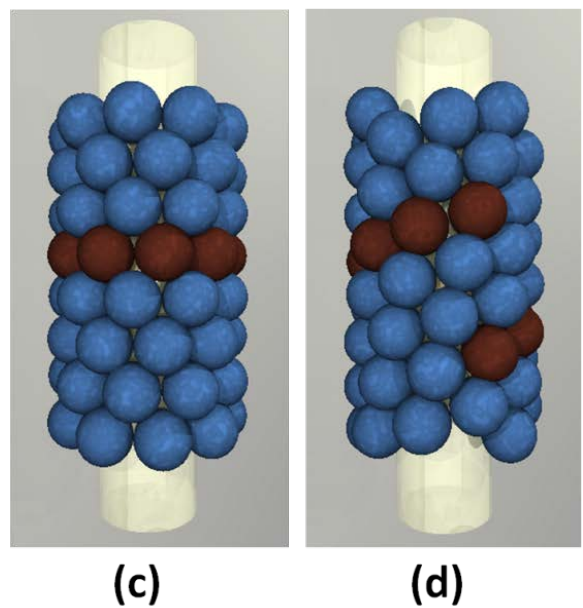

(a) (b) (c) (d)

**Fig. 8.** Two plausible stable configurations occur when the densely packed particle ring (top) is expanded and extra space is introduced: either (a) the extra space is shared among all the particles equally, or (b) the extra space is focused into a single pair of particles and the rest remain touching. (c) A single 'ring' of particles is highlighted in red/black. (d) A single helical coil of particles is highlighted in red.

$$E_{defect} = (N-1) \cdot V(d) + V(P - Nd + d),  \quad \text{(Eq. 8b)}$$

where *N* is the number of particles in the ring, *P* is the perimeter of the polyhedron connecting the centers of the particles (defined above and in Fig. 1c), and *d* is (as before) the location of the potential minima. When $E_{uniform}$ exceeds $E_{defect}$, each ring of particles on the cylinder will prefer to have a seam, and a uniform crystal lattice will no longer be stable. The critical *P* for which this transition occurs can be found by simply equating these two energies.

We can extend this approach to any value of θ by generalizing the ring to a segment of a helix that extends 2π along $\hat{\varphi}$ (Fig. 8d). This can be done approximately by replacing *P* with $P/\cos\theta_0$ and letting $N = P_0/(d\cos\theta_0)$, with $P_0$ and $\theta_0$ being the un-stretched values. Using these substitutions, we equate Eqs. 8a and 8b, and rearrange slightly to find

$$\left(\frac{P_0/d}{\cos\theta_0} - 1\right) \cdot V(d) = \frac{P_0/d}{\cos\theta_0} \cdot V\left(\frac{P}{P_0} d\right) - V\left(\frac{P}{\cos\theta_0} - \frac{P_0}{\cos\theta_0} + d\right), \quad \text{(Eq. 9)}$$

where

$$\cos\theta_0 = (n_1 + n_2/2)/\sqrt{n_1^2 + n_1 n_2 + n_2^2}. \quad \text{(Eq. 10)}$$

As the left side of Eq. 9 is a constant, the critical distance *P* can be solved computationally for any generic attractive potential *V(r)*, and the circumference *C* can be subsequently approximated from *P* via the relation

$$C \approx \pi d / \sin(\pi d / P). \quad \text{(Eq. 11)}$$

It is worth noting that, in the case of the Morse potential (Eq. 5), this critical value is determined by the parameter γd (which sets the range of the potential) in addition to $(n_1, n_2)$.

The results of these calculations are represented by the grey bars in the background of Figs. 4, 6, and 7. Calculated for all relevant combinations of $n_1$ and $n_2$, these bars each begin at $C_0$ (found via Eq. 2), and terminate at the *C* that marks the expected onset of line-slip defects. A comparison with the simulation data shows good agreement. When C/d is expanded beyond the ideal hexagonal value ($C_0$/d), in almost all cases the line-slip phases appear where Eq. 9 predicts. Additionally, much of the error in these predictions for low values of C/d is a product of the approximation used in Eq. 11; future work could improve the accuracy by instead numerically finding *C* from *P* using an exact trigonometric relation. When C/d is compressed *below* the perfect hexagonal value, however, the model predicts a line-slip phase for the Lennard-Jones potential, whereas simulations indicate that a crystal phase is stable under these conditions. This discrepancy is small when θ is close to 0° or 60°, but becomes significant when θ ≈ 30°. The source of the disagreement stems from a difficulty in applying our model to the type of transition in these regions; such a calculation requires an energetic comparison between particles in the line-slip phase and particles in the compressed oblique phase, which are two configurations with different values of *N* and θ. For this reason, correctly predicting the critical values of *C* that correspond to the onset of a compressed oblique lattice may require using a two-dimensional model instead.

Our model makes clear that as the range of attraction is decreased, the energy required to stretch the crystal lattice grows very quickly, and thus the line-slip phase is present within a greater percentage of the parameter space. As was stated previously, the extremely short-ranged γd = 30 Morse potential exists almost solely in the line-slip regime. We further infer that the steepness of the core repulsion in the chosen interaction potential governs where in parameter space the line-slip phases terminate; that is, the point where a compressed oblique lattice becomes favorable energetically to the line-slip phase. Once again, however, this transition is poorly described by our model.

Lastly, we note that the presence of crystalline and line-slip phases in parameter space can affect the dynamics of crystallization as the system is cooled. As an example, Fig. 9 shows images of one Lennard-Jones system undergoing freezing as the temperature is lowered. Here, the system first freezes into the $(n_1,n_2) = (6,0)$ crystal with $\theta = 0°$, but it is very highly stretched. As temperature is decreased, this crystal structure becomes unstable and is spontaneously replaced by the line-slip phase. At still lower temperatures, the line-slip transitions into a lower-energy crystalline configuration with $(n_1,n_2) = (6,1)$, as indicated in Fig. 4 with a star. This sequence of phases may be in accord with Ostwald's rule of stages, which postulates that crystallization proceeds with multiple metastable crystalline states in order of decreasing free energy barrier [30-32].

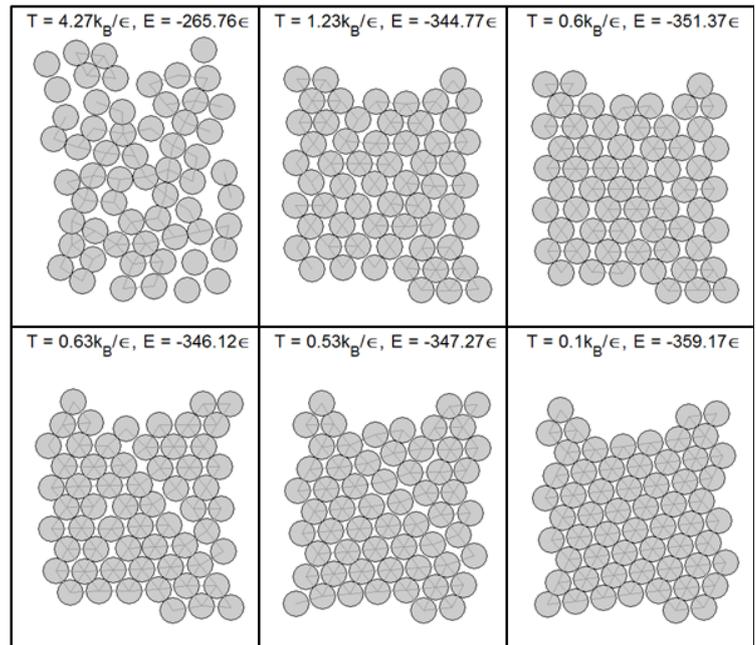

**Fig. 9.** Cooling of Lennard-Jones particles at $C/d$=6.76 (the location of which is indicated by a star in Fig. 4) into a stable crystal. The temperature and the energy, computed only from pairwise interactions, are indicated in each frame. Neighboring particles that are touching have been indicated with a bond. For clarity, bonds that cross the branch cut at $\varphi=2\pi$ have not been added. As T is lowered, we find a (6,0) phase, a line-slip phase, and finally the steady-state (6,1) phase.

## DISCUSSION

We have shown that, due to the constraints of the cylindrical geometry, at temperatures well below the planar-surface melting points (i.e. $k_BT \ll \epsilon$), an oblique crystal symmetry arises that is unstable on planar surfaces. We also find that a stable line-slip phase can form spontaneously within finite ranges of the cylinder circumference $C$, and that the regions in parameter space where line-slip phases are preferred broadens as the range of attraction is decreased. As direct evidence of this trend, our results indicate that the Lennard-Jones system favors the formation of oblique crystals, whereas the comparatively short ranged Morse potential with $\gamma d = 30$ almost exclusively forms line-slip structures. The primary features of this behavior are predicted by our straightforward one-dimensional model with reasonable accuracy. In particular, our results suggest that a line-slip phase can be energetically favorable for a variety of interaction potentials. It seems likely that more accurate analytical models of this problem could be developed, but the simplicity and transparency of our method allow it to be applied to other surface geometries and potentials. We can, perhaps, further extend the model to account for finite temperatures by including entropy. As a leading-order approximation, one could account for the free area of each sphere moving about its lattice site [33, 34]. In general, we expect hard spheres to embody the high-temperature limit, and the fact that the line-slip structure provides the maximum packing density when $C \neq C_0$ [16] implies that entropy may favor the line-slip. Hence, heating a crystal phase may lead to a line-slip phase unless pre-empted by melting.

In its current form, this model cannot be used to predict the behavior of purely repulsive or electrostatic interactions in this geometry without modification. One might assume, given the construction of Eq. 9, that a line-slip phase might never be energetically favorable for a repulsive potential. However, such a potential also has no inherent or preferred inter-particle spacing, and thus the optimal distance between the spheres should depend on the area fraction of spheres covering the cylinder surface. This complicates the problem since both $|a_1|$ and $|a_2|$ now vary, unlike the cases reported here where $|a_2|/d \approx 1$ throughout. In the purely repulsive case, the area fraction provides a constraint that again reduces the problem to a single dimension if the lattice structure is known.

Our results may have broad implications for understanding the way tubular crystals assemble in nature. In particular, it has been pointed out that many biological materials exhibit the type of structure described here, and therefore the same commensurability constraint [25, 26, 35]. Because capsid proteins in helical viruses (*tobacco mosaic virus*, for example) are constrained to bind to the surface of a RNA strand, the situation described here might be informative in understanding their structure. Microtubules also are capable of self-assembling with a seam–although in their case this seam is both purely longitudinal and achiral [36]. Nevertheless, such similarities are tantalizing enough to warrant further study. The straightforwardness of our predictions also suggest a practical application of these results as a novel method for producing crystalline media of desired symmetry and orientation on cylindrical surfaces. Specifically, one should be able to tune both the structure and orientation of a developing crystal lattice by adjusting the ratio C/d. Such crystals might spontaneously assemble more readily due to the existence of the line-slip phase, through which crystals that nucleate in a metastable structure can reorganize into a more stable configuration. Finally, we note that the bending rigidity of the crystal along the cylinder axis may depend on the structure. Bending a cylindrical crystal would lead to negative Gaussian curvature, and consequently to curvature-induced defects in the hexagonal and oblique crystal phases. By contrast, the 5-fold and 7-fold disclinations inherent in the line-slip phase might result in an overall bending rigidity that is different from that of a uniform crystal phase.

## ACKNOWLEDGMENTS

We thank Caitlin Nelson for her useful comments on this manuscript, and Ariel Amir for useful discussions. We also thank the National Science Foundation for funding through grant DMR-0907195 and through grant DMR-0846582.

## Notes and references

*Address, Department of Physics, University of Massachusetts Amherst, Hasbrouck Lab, 666 North Pleasant Street, Amherst, MA 01003, U.S.A. *E-mail:dinsmore@physics.umass.edu*

† Electronic Supplementary Information (ESI) available: explanation of the calculation of angles, θ. See DOI: 10.1039/b000000x/